\documentclass{article}

\usepackage[noadjust]{cite} 
\usepackage{arxiv}
\usepackage{cite}
\usepackage[utf8]{inputenc} 
\usepackage[T1]{fontenc}    
\usepackage{hyperref}       
\usepackage{url}            
\usepackage{booktabs}       
\usepackage{amsfonts}       
\usepackage{nicefrac}       
\usepackage[final,nopatch=footnote]{microtype}
\usepackage{lipsum}
\usepackage{graphicx}
\usepackage{subcaption}
\graphicspath{ {./images/} }
\usepackage{stmaryrd}
\usepackage{amsmath,amsthm}
\newtheorem*{definition}{Definition}
\newtheorem*{property}{Property}

\usepackage{setspace}
\usepackage{scalerel}
\usepackage{tikz}
\usetikzlibrary{svg.path}
\definecolor{orcidlogocol}{HTML}{A6CE39}
\usepackage{bm}
\usepackage{mathtools}
\usepackage{nicematrix}
\usepackage{mwe}

\newcommand{\myMatrix}[1]{\bm{\mathit{#1}}}

\tikzset{
	orcidlogo/.pic={
		\fill[orcidlogocol] svg{M256,128c0,70.7-57.3,128-128,128C57.3,256,0,198.7,0,128C0,57.3,57.3,0,128,0C198.7,0,256,57.3,256,128z};
		\fill[white] svg{M86.3,186.2H70.9V79.1h15.4v48.4V186.2z}
		svg{M108.9,79.1h41.6c39.6,0,57,28.3,57,53.6c0,27.5-21.5,53.6-56.8,53.6h-41.8V79.1z M124.3,172.4h24.5c34.9,0,42.9-26.5,42.9-39.7c0-21.5-13.7-39.7-43.7-39.7h-23.7V172.4z}
		svg{M88.7,56.8c0,5.5-4.5,10.1-10.1,10.1c-5.6,0-10.1-4.6-10.1-10.1c0-5.6,4.5-10.1,10.1-10.1C84.2,46.7,88.7,51.3,88.7,56.8z};
	}
}
\newcommand\orcidicon[1]{\href{https://orcid.org/#1}{\mbox{\scalerel*{
				\begin{tikzpicture}[yscale=-1,transform shape]
					\pic{orcidlogo};
				\end{tikzpicture}
			}{|}}}}
\usepackage{hyperref}

\title{Circulating Currents in Electric Machines:

Positive Impact of The End Windings Length on Losses}

\author{
 Taha El Hajji \orcidicon{0000-0002-0168-8180}\\
  Department of Electrical Engineering and Automation\\
  Aalto University\\
  P.O. Box 15500, 00076 Espoo, Finland\\
  \texttt{taha.elhajji@aalto.fi} \\
   \And
 Antti Lehikoinen \orcidicon{0000-0001-8170-4146}\\
  Department of Electrical Engineering and Automation\\
  Aalto University\\
  P.O. Box 15500, 00076 Espoo, Finland\\
  Smeklab Ltd, Espoo, Finland\\
  \texttt{antti.lehikoinen@aalto.fi} \\
  \And
 Anouar Belahcen \orcidicon{0000-0003-2154-8692}\\
  Department of Electrical Engineering and Automation\\
  Aalto University\\
  P.O. Box 15500, 00076 Espoo, Finland\\
  \texttt{anouar.belahcen@aalto.fi} \\
}

\begin{document}

\maketitle
\begin{abstract}
Circulating currents occurring in windings of electric machines received rising interest recent years. Circulating currents represent unwanted currents flowing between parallel-connected conductors. This phenomenon is due to various reasons such as asymmetries in the winding and differences in electric potential between parallel-connected conductors. This effect occurs both at no-load and on-load conditions, and always lead to uneven distribution of the current between the parallel conductors, therefore leading to higher losses, as proven in the authors' previous work. Circulating currents are occurring mainly due to asymmetries and electric potential difference in the active part, meaning that long end windings are advantageous to mitigate the effect of circulating currents. Losses due to circulating currents decrease at a rate proportional to the inverse square of the end windings length. The aim of this paper is to mathematically prove this property and present a case study application in an electric machine.

\end{abstract}


\section{Introduction}

The phonemenon of circulating currents is defined and reviewed in \cite{arxiv_taha_2024_1,Access_Taha_2024}, with an application in electric machines. Circulating currents become a concern in the windings of electric machines especially at high rotational speeds \cite{Machines_Taha_2023,Aerospace_Taha_2024,HighSpeed_1CN,ICEM_Taha_2024,PhD_Taha_2023}. The same phenomenon is observed in transformers and inductors operating at high frequencies \cite{Transformer_1,Transformer_2,Transformer_4,Transformer_5,Transformer_6,Transformer_7}. Circulating currents cause uneven distribution of the total current, flowing in the bundle, between the parallel-connected conductors. This occurs also at no load, meaning that circulating currents are flowing between the parallel conductors, with a total current being zero. This uneven distribution always leads to higher losses in the windings compared to the case of even distribution of current between the conductors. This fundamental property of circulating currents is proved mathematically by authors in \cite{arxiv_taha_2024_1} with a case application in electric machines. The causes of circulating currents and the drawbacks are also discussed in \cite{arxiv_taha_2024_1}. The major causes are the differences in the inductance and the electric potential, which can be due to different placements of parallel conductors inside the slots \cite{Placement_Random_1_RefFI_Antti,Placement_Random_2_RefFI_Antti,Placement_Random_3_PhD_Antti_2017,Placement_Precise_2_RefFI_PhD_Jahirul_2010,Asymmetry_1,Abnormality_1}. Circulating currents occur also in other situations in electric machines \cite{pramodCC}, however, in this work, we focus on this effect occurring in parallel-connected conductors. There exists other phenomena in windings, not covered in this paper, such as the skin effect \cite{Access_Taha_2024,ICEM_Taha_2020} and the proximity effect \cite{Access_Taha_2024,ICEM_Taha_2020,SkinProx_1_UKSE,SkinProx_2_USA_Ayman,SkinProx_3_DE,SkinProx_4_UK}, causing higher losses. These effects have been previously reviewed by authors in \cite{Access_Taha_2024}, summarizing the main models in the literature, with an application in electric machines.

Several solutions have been proposed in the literature to reduce losses due to circulating currents such as transposition \cite{Transposition_1_Ref22,Transposition_2_Ref23,Transposition_3_RefCN1,Transposition_4_RefCN2,Transposition_5_RefNW,Transposition_6_RefUK,Transposition_7_RefJP,Transposition_8_RefCN}.
Circulating currents can also be mitigated by increasing the length of the end windings. This will result in higher joules losses, however, it might be beneficial in situations where circulating currents losses are very high. More specifically, losses due to circulating currents decrease at a rate proportional to the inverse square of the end windings length. This property remains valid for conductors with both circular \cite{Circular_Wire_Ref18,Circular_Wire_Ref19,Circular_Wire_Ref20} and rectangular shapes \cite{Guideline_1,Guideline_2,Rectangular_Wire_1}. The aim of this paper is to present a proof of this this property in a mathematical framework.

The main models, allowing the evaluation of losses due circulating currents in electric machine's windings are reviewed in \cite{Access_Taha_2024} and are categorized into three main types: finite element models coupled with circuit analysis \cite{FEA_1_Antti}, analytical models \cite{Placement_Random_1_RefFI_Antti,Transposition_1_Ref22,Transposition_2_Ref23}, and hybrid models \cite{Hybrid_Ref33}, combining results from finite element analysis and analytical formulas. The hybrid model \cite{Hybrid_Ref33} is used in this work to prove the property mentioned above.

The basic terms and the preliminary formulas used in this paper are listed in Section 2. The definition of circulating currents \cite{arxiv_taha_2024_1} is re-presented in Section 3 as well as the model used. Section 4, presents the property mentioned above, initially discussed in previous authors' work\footnotemark{} (Appendix A of \cite{Access_Taha_2024}), but extended in this paper in a detailed and complete version, along with a mathematical proof framework and a case application in electric machines in Section 5.

\footnotetext{The property and its proof can be referred to using this paper and the preceding paper \cite{Access_Taha_2024}: \url{https://ieeexplore.ieee.org/document/10366258}.}

\section{Terminology and Preliminary formulas}

The terminology related to windings and circulating currents were previously listed in \cite{arxiv_taha_2024_1}. This terminology is reminded below and the preliminary formulas \cite{Access_Taha_2024,Hybrid_Ref33} are also listed.

\begin{itemize}
    \item \textbf{Strand}: A single conductor often with small dimensions (to reduce losses due to the skin and proximity effects).
    \item \textbf{Bundle}: A winding composed of multiple strands connected in parallel ($Nsh$ is referred to in the following as the total number of parallel strands).
    \item \textbf{Bundle-level proximity effect losses} or \textbf{Circulating currents losses}: Overall Losses in the bundle due to circulating currents.
    \item \textbf{$T$}: Fundamental period of the AC electrical quantities.
    \item \textbf{$\omega$}: Fundamental pulsation of the AC electrical quantities.
    \item \textbf{$N_{h}$}: Number of harmonics.
    \item \textbf{$I_{bundle}(t)$}: Instantaneous total current flowing in the bundle.
    \item \textbf{$I_i (t)$}: Instantaneous current flowing in the strand $i$ considering the circulating currents.
    \item \textbf{$I_{RMS}$}: RMS value of the bundle's total current:
    \begin{equation}
    I_{RMS} = \sqrt{\frac{1}{T} \int_{0}^{T}I_{bundle}^2(t) \,dt} \\
\label{RMSexpression1}
\end{equation}
    \item \textbf{$I_{RMS,i}$}: RMS value of the current in the strand $i$ considering the effect of circulating currents:
    \begin{equation}
    I_{RMS,i} = \sqrt{\frac{1}{T} \int_{0}^{T}I_i^2(t) \,dt} \\
\label{RMSexpression2}
\end{equation}
    \item \textbf{$I_{bundle,k}$}: RMS value of the harmonic $k$ of the total current flowing in the bundle.
    \item \textbf{$I_{i,k}$}: RMS value of the harmonic $k$ of the current in the strand $i$ considering the circulating currents.
    \item \textbf{$r_{strd}$}: Radius of the strand (in the case of circular shape).
    \item \textbf{$l_{EW}$}: End winding length of the machine (illustrated in Figure \ref{alpha}).
    \item \textbf{$l_{active}$}: Active length of the machine (illustrated in Figure \ref{alpha}).
    \item \textbf{$l_{coil}$}: Length of one coil turn:
    \begin{equation}
        l_{coil} = 2\times (l_{active} + l_{EW})
    \end{equation}
    \item \textbf{$\alpha_w$}: Ratio of half coil length over the active length:
    \begin{equation}
    \alpha_w = \frac{l_{coil}}{2 \times l_{active}}
    \label{equ_alpha}
    \end{equation}
    \item \textbf{$P_{CC=0,i}$}: DC Losses occurring in the strand $i$ without the effect of circulating currents.
    \item \textbf{$P_{CC=0}$}: DC Losses occuring in the bundle without the effect of circulating currents:
    \begin{equation}
    P_{CC=0}= \sum_{i=1}^{Nsh} P_{CC=0,i} \\
\end{equation}
    \item \textbf{$P_{CC,i}$}: Circulating currents losses in the strand $i$.
    \item \textbf{$P_{CC}$}: Circulating currents losses in the bundle:
    \begin{equation}
    P_{CC}= \sum_{i=1}^{Nsh} P_{CC,i} \\
\end{equation}
    \item $P_{\delta CC}$: Additional losses due to the effect of circulating currents:
    \begin{equation}
        P_{\delta CC} = P_{CC} - P_{CC=0}
    \end{equation}
    \item $P_{\delta CC, active}$: Additional losses due to the effect of circulating currents if only the active part of the machine is considered and the end windings are neglected ($l_{EW}=0$).
    \item \textbf{$N_{slots}$}: Number of slots in the machine's stator.
    \item \textbf{$p$}: Number of poles in the machine's rotor.
    \item \textbf{$N_c$}: Number of conductors in one slot.
    \item \textbf{$S_i$}: Slot $i$ of the machine's stator.
    \item \textbf{$N_{s,A}$}: Number of the slots that contain the conductors of the phase A.
    \item \textbf{$N_{t-A,i}$}: Number of turns of the phase A in the slot $i$ of the machine's stator.
    \item \textbf{$Nsh_{A,i}$}: Number of the parallel strands of the phase A in the slot $i$.
    \item \textbf{$N_{p-s}$}: Number of the parallel-connected slots containing conductors of the phase A.
    \item \textbf{$Nsh$}: Number of strands in hand,  which corresponds in our case to the number of parallel strands in all the machine's stator for each phase:
    \begin{equation}
    N_{sh} = N_{p-s} \times Nsh_{A,i} \\
\end{equation}
    \item \textbf{$SPP$}: Number of slots per phase and per pole.
    \item \textbf{$L_{strd}$}: Self inductance of each strand.
    \item \textbf{$R_{strd}$}: Resistance of each strand.
    \item \textbf{$Z_{strd,k}$}: Self impedance of each strand for the harmonic $k$:
    \begin{equation}
        Z_{strd,k} = R_{strd} + j.k.\omega.L_{strd}
    \end{equation}
    \item \textbf{$L_{strd,act}$}: Self inductance of each strand considering only the active part of the machine.
    \item \textbf{$R_{strd,act}$}: Resistance of each strand considering only the active part of the machine.
    \item \textbf{$Z_{strd,act,k}$}: Self impedance of each strand for the harmonic $k$ considering only the active part of the machine:
    \begin{equation}
        Z_{strd,act,k} = \frac{1}{\alpha _w} . Z_{strd,k}
    \end{equation}
    \item \textbf{$\myMatrix{I_n} \in \mathbb{R}^{n \times n}$}: Identity matrix with $n$ rows and $n$ columns.
    \item \textbf{$\myMatrix{J_{n,m}} \in \mathbb{R}^{n \times m}$}: Matrix of ones with $n$ rows and $m$ columns (all coefficients are equal to $1$).
    \item \textbf{$Cd_k$}: Conductor $k$ in a slot.
    \item \textbf{$Str_l$}: Strand $l$ in a slot.
    \item \textbf{$Dist(i,j)$}: Distance between conductors $Cd_i$ and $Cd_j$.
    \item \textbf{$\Delta \varphi$}: Potential difference of the parallel strands.
    \item \textbf{$\Delta \varphi _{k}$}: Potential difference of the parallel strands for the harmonic $k$.
    \item \textbf{$\myMatrix{\Delta} \varphi \in \mathbb{R}^{Nsh \times 1}$}: Potential difference vector:
    \begin{equation}
    \begin{aligned}
    \myMatrix{\Delta} \varphi &=  \Delta \varphi . \begin{bmatrix}
           1 \\
           \vdots \\
           1
         \end{bmatrix}\\
    \end{aligned}
    \end{equation}
    \item \textbf{$\myMatrix{A_{z,e,i}} \in \mathbb{R}^{N_c \times 1}$}: $A_{z,e,i}(l)$ is the external vector potential at the location of the conductor $l$ in the slot $i$.
    \item \textbf{$\myMatrix{M_{A-i}} \in \mathbb{R}^{N_c \times Nsh}$}: Winding matrix of the phase A in slot $i$:
    \begin{equation*}
    \begin{aligned}
    \myMatrix{M_{A-i}} (k,l) & =
    \begin{cases}
    \begin{aligned}
    1, & \textnormal{  } Cd_k \textnormal{ linked to } Sr_l \textnormal{ in + coil side in the slot } S_i \\      
    -1, & \textnormal{  } Cd_k \textnormal{ linked to } Sr_l \textnormal{ in - coil side in the slot } S_i \\
    0, &  \textnormal{  } Cd_k \textnormal{ not connected to } Sr_l \textnormal{ in the slot } S_i 
    \end{aligned}
    \end{cases}
    \end{aligned}
    \end{equation*}
    \item \textbf{$\myMatrix{R_{i}} \in \mathbb{R}^{N_c \times N_c}$}: Resistance matrix of conductors in the slot $i$:
    \begin{equation*}
    \begin{aligned}
    \myMatrix{R_{i}} (k,l) & =
    \begin{cases}
    \begin{aligned}
    & \frac{l_{cond}}{\sigma \times S}, & \textnormal{  } k = l \\
    & 0, & \textnormal{  } k \neq l
    \end{aligned}
    \end{cases}
    \end{aligned}
    \end{equation*}
    \item \textbf{$\myMatrix{L_{i}} \in \mathbb{R}^{N_c \times N_c}$}: Inductance matrix of conductors in the slot $i$:
    \begin{equation*}
    \begin{aligned}
    \myMatrix{L_{i}} (k,l) & =
    \begin{cases}
    \begin{aligned}
    &- \frac{\mu_{0} \times l_{cond}}{8 \times \pi}, & \textnormal{  } k = l \\
    &- \frac{\mu_{0} \times l_{cond}}{4 \times \pi}.\left[ \ln\left[ \frac{Dist^2(k,l)}{r_{strd}^2} \right] +1\right], & \textnormal{  } k \neq l
    \end{aligned}
    \end{cases}
    \end{aligned}
    \end{equation*}
    \item \textbf{$\myMatrix{\phi _i} \in \mathbb{R}^{N_c \times 1}$}: External flux vector at the conductors in the slot $i$:
    \begin{equation}
        \myMatrix{\phi _i} = l_{active} \times \myMatrix{A_{z,e,i}}
    \end{equation}
    \item \textbf{$\myMatrix{\phi _{i,k}} \in \mathbb{R}^{N_c \times 1}$}: Harmonic $k$ of the external flux vector in the slot $i$.
    \item \textbf{$\myMatrix{R_{A-i}} \in \mathbb{R}^{Nsh \times Nsh}$}: Resistance Matrix of the phase A in the slot $i$:
    \begin{equation}
        \myMatrix{R_{A-i}} = \myMatrix{M_{A-i}^T} \times \myMatrix{R_{i}} \times \myMatrix{M_{A-i}}
    \end{equation}
    \item \textbf{$\myMatrix{L_{A-i}} \in \mathbb{R}^{Nsh \times Nsh}$}: Inductance Matrix of the phase A in the slot $i$:
    \begin{equation}
        \myMatrix{L_{A-i}} = \myMatrix{M_{A-i}^T} \times \myMatrix{L_{i}} \times \myMatrix{M_{A-i}}
    \end{equation}
    \item \textbf{$\myMatrix{\phi _{A-i}} \in \mathbb{R}^{Nsh \times Nsh}$}: External flux vector at the conductors of the phase A in the slot $i$:
    \begin{equation}
        \myMatrix{\phi _{A-i}} = \myMatrix{M_{A-i}^T} \times \myMatrix{\phi _{i}}
    \end{equation}
    \item \textbf{$\myMatrix{\phi _{A-i,k}} \in \mathbb{R}^{Nsh \times Nsh}$}: Harmonic $k$ of the external flux vector at the conductors of the phase A in the slot $i$.
    \item \textbf{$\myMatrix{R_{A}} \in \mathbb{R}^{Nsh \times Nsh}$}: Resistance Matrix of the phase A:
    \begin{equation}
        \myMatrix{R_{A}} = \sum_{i=1}^{N_{s,A}} \myMatrix{R_{A-i}}
    \end{equation}
    \item \textbf{$\myMatrix{L_{A}} \in \mathbb{R}^{Nsh \times Nsh}$}: Inductance Matrix of the phase A:
    \begin{equation}
        \myMatrix{L_{A}} = \sum_{i=1}^{N_{s,A}} \myMatrix{L_{A-i}}
    \end{equation}
    \item \textbf{$\myMatrix{\phi _{A}} \in \mathbb{R}^{Nsh \times Nsh}$}: External flux vector at the conductors of the phase A:
    \begin{equation}
        \myMatrix{\phi_{A}} = \sum_{i=1}^{N_{s,A}}\myMatrix{\phi_{A-i}}
    \end{equation}
    \item \textbf{$\myMatrix{\phi _{A,k}} \in \mathbb{R}^{Nsh \times Nsh}$}: Harmonic $k$ of the external flux vector at the conductors of the phase A.
    \item \textbf{$\myMatrix{Z_{A,k}} \in \mathbb{R}^{Nsh \times Nsh}$}: Impedance Matrix of the phase A for the harmonic $k$:
    \begin{equation}
    \myMatrix{Z_{A,k}} = \myMatrix{R_A} + j.k.\omega.\myMatrix{L_A}
    \end{equation}
    \item \textbf{$\myMatrix{A_{k}} \in \mathbb{R}^{(Nsh+1) \times (Nsh+1)}$}: System Matrix for the harmonic $k$:
    \begin{equation}
    \myMatrix{A_k} = \left[\begin{array}{cc}
    \myMatrix{Z_{A,k}} & \begin{array}{c}
    1 \\
    \vdots \\
    1\\ \end{array} \\
    \begin{array}{ccc}
    1 & \cdots & 1
    \end{array} & 0 \end{array}\right]
    \end{equation}
    \item \textbf{$\myMatrix{X_{k}} \in \mathbb{R}^{(Nsh+1) \times 1}$}: Unknown vector for the harmonic $k$:
    \begin{equation}
    \myMatrix{X_k} = \left[\begin{array}{c}
    I_{1,k} \\
    \vdots \\
    I_{Nsh,k} \\ 
    \Delta \varphi _{k}\\ \end{array}\right]
    \label{Xk}
    \end{equation}
    \item \textbf{$\myMatrix{B_{k}} \in \mathbb{R}^{(Nsh+1) \times 1}$}: Output vector for the harmonic $k$:
    \begin{equation}
    \myMatrix{B_k} = \left[\begin{array}{c}
    -j.(k.\omega).\myMatrix{\phi _{A,k}} \\ 
    I_{bundle,k}\\ \end{array}\right]
    \end{equation}
    \item \textbf{$\myMatrix{A} \in \mathbb{R}^{((Nsh+1).N_h) \times ((Nsh+1).N_h)}$}: System Matrix for all the harmonics:
    \begin{equation}
    \myMatrix{A} = \left[\begin{array}{ccc}
    \myMatrix{A_1} & \cdots & 0 \\
    \vdots & \ddots & \vdots \\
    0 & \cdots & \myMatrix{A_{N_h}} \\
    \end{array}\right]
    \end{equation}
    \item \textbf{$\myMatrix{X} \in \mathbb{R}^{((Nsh+1).N_h) \times 1}$}: Unknown vector for all the harmonics:
    \begin{equation}
    \myMatrix{X} = \left[\begin{array}{c}
    \myMatrix{X_1} \\
    \vdots \\
    \myMatrix{X_{N_h}} \\ \end{array}\right]
    \end{equation}
    \item \textbf{$\myMatrix{B} \in \mathbb{R}^{((Nsh+1).N_h) \times 1}$}: Output vector for all the harmonics:
    \begin{equation}
    \myMatrix{B} = \left[\begin{array}{c}
    \myMatrix{B_1} \\ 
    \vdots \\
    \myMatrix{B_{N_h}} \\ \end{array}\right]
    \end{equation}
\end{itemize}

\begin{figure}[htbp]
\begin{center}
\includegraphics[scale=0.5]{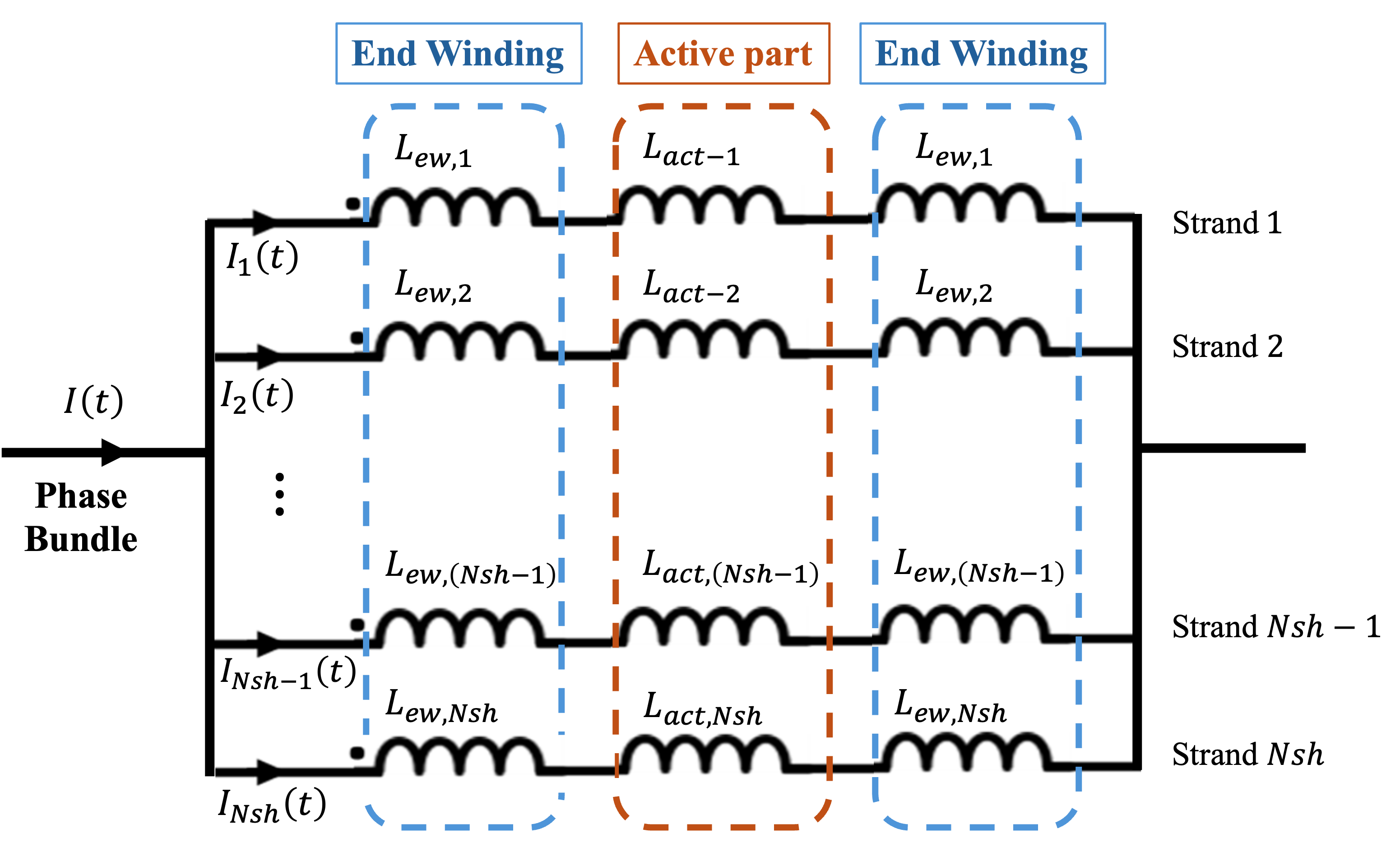}
\caption{Phase bundle composed of $Nsh$ parallel strands}
\label{CC_EW}
\end{center}
\end{figure}

In this paper, we focus on the AC supply, however, the results can be generalized to the DC case.
Figure \ref{CC_EW} shows the case of $n$ parallel-connected strands of an electric machine's phase, all having the same dimensions and, therefore, similar DC resistance.

\section{Circulating Currents: Mathematical Definition and Model}

Circulating currents have been previously defined in \cite{arxiv_taha_2024_1} in a mathematical form. The definition of circulating currents is:

\begin{definition}
    Circulating currents \underline{occur} between $Nsh$ parallel-connected strands, if and only if:
    \begin{equation}
        \exists i \in \llbracket 1,Nsh\rrbracket, \exists t \in [0,T],  I_i (t) \neq \frac{I_{bundle}(t)}{Nsh}
    \label{definition1}
    \end{equation}
\end{definition}

Therefore, the definition of no circulating currents occurring between parallel-connected strands is:
\begin{definition}
    Circulating currents \underline{do not occur} between $Nsh$ parallel-connected strands, if and only if:
    \begin{equation}
        \forall i \in \llbracket 1,Nsh\rrbracket, \forall t \in [0,T],  I_i (t) = \frac{I_{bundle}(t)}{Nsh}
    \label{definition2}
    \end{equation}
\end{definition}

\textbf{Model for Circulating Currents:}

Among the various existing models in the literature \cite{Access_Taha_2024} to evaluate losses due circulating currents (finite element analysis coupled with circuit analysis, analytical models, and hybrid models), we are using in this work the hybrid model proposed in \cite{Hybrid_Ref33}. This hybrid model allows evaluating circulating currents flowing in the parallel-connected strands by combining results from finite element analysis (potential vector value at each conductor) and analytical formulas. The matrix equation to be solved is the following:

\begin{equation}
    \myMatrix{A} . \myMatrix{X} = \myMatrix{B}
    \label{HM}
\end{equation}

Where the the matrix $\myMatrix{A}$ and the vectors $\myMatrix{X}$ and $\myMatrix{B}$ are expressed in Section 2.
The vector $\myMatrix{B}$ is obtained based on the predefined value of the total current flowing in the bundle $I_{bundle}$ and the external flux, obtained by evaluating the external vector potential $\myMatrix{A_{z,e,i}}$ at each conductors with finite element analysis. The solution of (\ref{HM}) gives the current flowing in each strand.

\section{Positive Impact of End Windings on Circulating Currents Losses: Property and Proof}

\begin{figure}[ht]
\begin{center}
\includegraphics[scale=0.4]{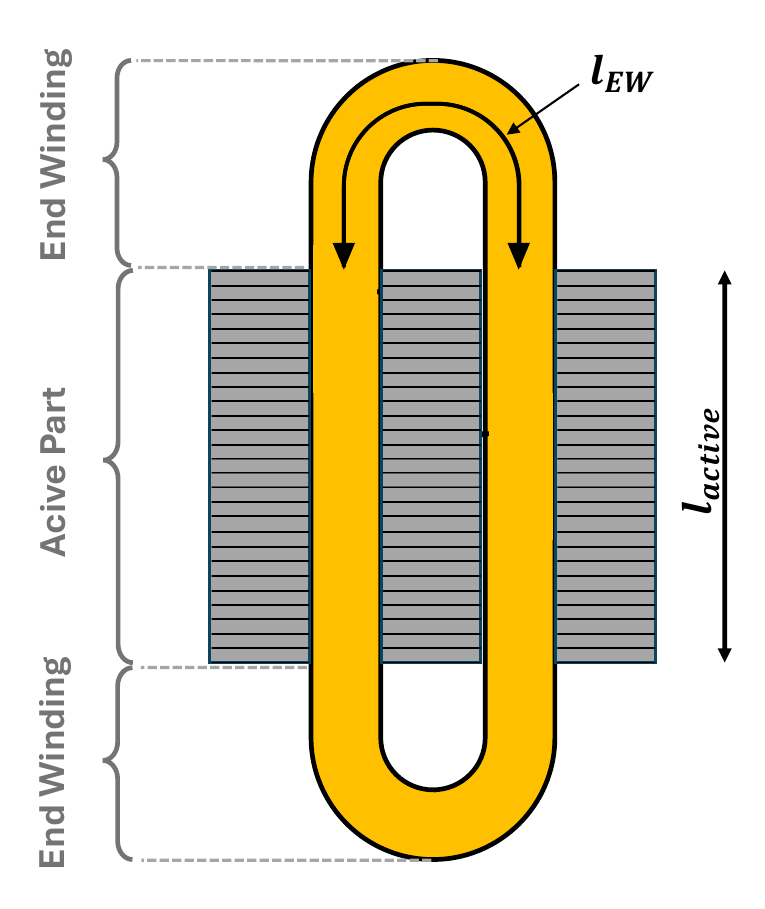}
\caption{Illustration of the ratio $\alpha_w$}
\label{alpha}
\end{center}
\end{figure}

The ratio $\alpha_w$ (\ref{equ_alpha}), used in the enunciation and the proof of the property below, can also be expressed using the dimensions illustrated in Figure \ref{alpha} as:
\begin{equation*}
    \alpha_w = 1+ \frac{l_{EW}}{l_{active}}
\end{equation*}

\begin{property}\footnotemark{}
\begin{center}
\begin{tabular}{rl}
    The ratio $\alpha_w$ increases $\Rightarrow$ & The losses due to circulating currents $P_{CC}$ decrease at a rate proportional to $\frac{1}{\alpha_w^2}$  \\
    & with the lower limit being $P_{CC=0}$, the losses occurring without the circulating currents
\end{tabular}
\end{center}
\label{property}
\end{property}

\footnotetext{The property, initially discussed in \cite{Access_Taha_2024}, is extended in this work and proved in a mathematical context. This work can be cited using this paper and the preceding paper \cite{Access_Taha_2024}: \url{https://ieeexplore.ieee.org/document/10366258}.}

This property means that if the length of the end windings increases compared to the active length of the machine, then the losses due to circulating currents decrease at a rate proportional to the inverse square of $\alpha_w$, and therefore, proportional to the inverse square of the end windings length. This property was partially enunciated and partially demonstrated by authors in \cite{Access_Taha_2024} and is extended here to a complete form with a full mathematical proof.

\begin{proof}

The property's proof is given in these two cases:
\begin{enumerate}
    \item The inductance matrix $\myMatrix{L_A}$ is negligible compared to the resistance matrix $\myMatrix{R_A}$ of the phase A: $\myMatrix{L_A}<<\myMatrix{R_A}$.
    \item The mutual inductance between conductors is negligible compared to the self inductance of conductors, therefore, the inductance matrix $\myMatrix{L_A}$ can be considered a diagonal matrix.
\end{enumerate}

The impedance matrix can be considered in these two cases as a diagonal matrix. Therefore, let's assume that the impedance matrix $\myMatrix{Z_{A,k}}$ is diagonal, which can be expressed as:
\begin{equation}
    \myMatrix{Z_{A,k}} = Z_{strd,k}.\myMatrix{I_{Nsh}}
\end{equation}

Which can be expressed also as:
\begin{equation}
    \myMatrix{Z_{A,k}} = \alpha_w . Z_{strd,act,k}.\myMatrix{I_{Nsh}}
\end{equation}

The system matrix $\myMatrix{A_k}$ for the harmonic $k$ can then be expressed as follows:
\renewcommand{\arraystretch}{1.5}
\begin{equation}
    \myMatrix{A_k} = \left[\begin{array}{ccc|c}
    (\alpha_w . Z_{strd,act,k}) & \cdots & 0 & 1 \\
    \vdots & \ddots & \vdots & \vdots \\
    0 & \cdots & (\alpha_w . Z_{strd,act,k}) & 1\\
    \hline
    1 & \cdots & 1 & 0
    \end{array} \right]
\end{equation}

The inverse of this matrix $\myMatrix{A_k}$ can be expressed as:
\begin{equation}
    \myMatrix{A_k}^{-1} = \left[\begin{array}{c|c}
    \myMatrix{\frac{1}{\alpha_w}.P_k}  & \begin{array}{c}
    \frac{1}{Nsh}      \\
    \vdots \\
    \frac{1}{Nsh}
    \end{array} \\
    \hline
    \begin{array}{ccc}
    \frac{1}{Nsh} & \cdots & \frac{1}{Nsh}
    \end{array} & \frac{-(\alpha_w . Z_{strd,act,k})}{Nsh}
    \end{array}\right]
\end{equation}

Where the matrix $\myMatrix{P_k} \in \mathbb{R}^{Nsh \times Nsh}$ is expressed as:
\begin{equation}
    \myMatrix{P_k} = \frac{1}{Z_{strd,act,k}}.\left[\myMatrix{I_{Nsh}}-\frac{1}{Nsh}.\myMatrix{J_{Nsh,Nsh}}\right]
\end{equation}

Hence, the inverse of the matrix $\myMatrix{A}$ can be expressed as:
\begin{equation}
    \myMatrix{A^{-1}} = \left[ \begin{array}{ccc}
    \myMatrix{A_1^{-1}} & \cdots & 0 \\
    \vdots & \ddots & \vdots \\
    0  &  \cdots & \myMatrix{A_{N_h}^{-1}} 
    \end{array}\right]
\end{equation}

The unknown vector $\myMatrix{X}$ can be obtained using the following equation:
\begin{equation}
    \myMatrix{X} = \myMatrix{A^{-1}}.\myMatrix{B}
\end{equation}

Since the operation is done by block, the unknown vector $\myMatrix{X_k}$ for the harmonic $k$ can be obtained similarly as:
\begin{equation}
    \myMatrix{X_k} = \myMatrix{A_k^{-1}}.\myMatrix{B_k}
\end{equation}

which can be formulated as:
\begin{equation}
    \myMatrix{X_k} = \left[ \begin{array}{c}
         -j.k.\omega.\frac{1}{\alpha_w}.\myMatrix{P_k}.\myMatrix{\phi_{A,k}} + \frac{1}{Nsh}.I_{bundle,k}.\myMatrix{J_{Nsh,1}}  \\
          -j.k.\omega.\frac{1}{Nsh}.\myMatrix{J_{1,Nsh}}.\myMatrix{\phi_{A,k}} -\frac{\alpha_w.Z_{strd,act,k}}{Nsh}.I_{bundle,k}
    \end{array}
    \right]
\end{equation}

Based on the expression of the vector $\myMatrix{X_k}$ in (\ref{Xk}), we have:
\begin{equation}
    \left[ \begin{array}{c}
     I_{1,k} \\
    \vdots \\
    I_{Nsh,k} \end{array} \right] = -j.k.\omega.\frac{1}{\alpha_w}.\myMatrix{P_k}.\myMatrix{\phi_{A,k}} + \frac{1}{Nsh}.I_{bundle,k}.\myMatrix{J_{Nsh,1}}
\end{equation}
Which can also be written as:
\begin{equation}
    \left[ \begin{array}{c}
     I_{1,k} \\
    \vdots \\
    I_{Nsh,k} \end{array} \right] = \left[ \begin{array}{c}
     \frac{I_{bundle,k}}{Nsh} \\
    \vdots \\
    \frac{I_{bundle,k}}{Nsh} \end{array} \right] + \frac{1}{\alpha_w} .\left[ \begin{array}{c}
     I_{\delta,1,k} \\
    \vdots \\
    I_{\delta,Nsh,k}
    \end{array} \right]
\end{equation}

The current in each strand can be expressed as:

\begin{equation}
    \forall j \in \llbracket 1,Nsh\rrbracket, \forall k \in \llbracket 1,N_h\rrbracket, I_{j,k} = \frac{I_{bundle,k}}{Nsh} + \frac{1}{\alpha_w}. I_{\delta,j,k}
    \label{I_strand}
\end{equation}

Where, for a strand $j \in \llbracket 1,Nsh\rrbracket$ and a harmonic $k \in \llbracket 1,N_h\rrbracket$, we have:
\begin{equation}
    I_{\delta,j,k} = \frac{-j.k.\omega}{Z_{strd,act,k}}.\left[ \myMatrix{\phi_{A,k}}(j)-\frac{1}{Nsh}.\sum_{n=1}^{Nsh}\myMatrix{\phi_{A,k}}(n)\right]
    \label{I_strand_component}
\end{equation}

It can be noticed from (\ref{I_strand}) that the current in each strand is composed of two terms. The first term $\frac{I_{bundle,k}}{Nsh}$ corresponds to the uniform sharing of the bundle's current between parallel strands. The second term $\frac{1}{\alpha_w}.I_{\delta,j,k}$ represents the effect of circulating currents. Therefore, when circulating currents are not considered, only the first term remains, meaning that the current flowing in the bundle is evenly shared between the $Nsh$ parallel-connected strands. Further, if the ratio $\alpha_w$ increases, corresponding to increasing the length of the end windings compared to the active length, the term $\frac{1}{\alpha_w}.I_{\delta,j,k}$ tends to $0$, and therefore, only the term $\frac{I_{bundle,k}}{Nsh}$ remains. This means that the current tends to be equally shared between parallel-connected strands. Hence, we obtain in this case:

\begin{equation}
    \left[ \begin{array}{c}
     I_{1,k} \\
    \vdots \\
    I_{Nsh,k} \end{array} \right] = \left[ \begin{array}{c}
     \frac{I_{bundle,k}}{Nsh} \\
    \vdots \\
    \frac{I_{bundle,k}}{Nsh} \end{array} \right]
\end{equation}

The RMS value of the current in each strand $j$ can expressed, according to Parseval equality, as:
\begin{equation}
    I_{RMS,j} = \sqrt{\sum_{k=1}^{N_h} I_{j,k}^2}
\end{equation}

Therefore, losses in each strand $j \in \llbracket 1,Nsh\rrbracket$ are expressed as:
\begin{equation}
    P_{CC,j} = R_{strd} \times I_{RMS,j}^2 = R_{strd} \times \sum_{k=1}^{N_h} I_{j,k}^2
\end{equation}

Total Losses in the bundle can then be expressed as follows:
\begin{equation}
    \begin{aligned}
        P_{CC} & =  \sum_{j=1}^{Nsh} P{_{CC,j}} \\
        & = R_{strd} \times \sum_{j=1}^{Nsh} \left[ \sum_{k=1}^{N_h} I_{j,k}^2 \right] \\
        & = R_{strd} \times \sum_{k=1}^{N_h} \left[
        \sum_{j=1}^{Nsh} I_{j,k}^2 \right]
    \end{aligned}
\end{equation}

Which, after substituting with (\ref{I_strand}), yields:
\begin{equation}
    \begin{aligned}
        P_{CC}  = {} &R_{strd} \times \sum_{k=1}^{N_h} \left[
        \sum_{j=1}^{Nsh} \left( \frac{I_{bundle,k}}{Nsh} + \frac{1}{\alpha_w}.I_{\delta,j,k} \right)^2 \right] \\
        = {} & R_{strd} \times \sum_{k=1}^{N_h} \left[
        \sum_{j=1}^{Nsh} \left[ \left(\frac{I_{bundle,k}}{Nsh} \right)^2 + 2 \times \frac{I_{bundle,k}}{\alpha_w.Nsh} . I_{\delta,j,k} + \left( \frac{1}{\alpha_w}.I_{\delta,j,k} \right)^2 \right] \right] \\
        = {} & \underbrace{ R_{strd} \times \sum_{k=1}^{N_h}
        \sum_{j=1}^{Nsh} \left[ \left(\frac{I_{bundle,k}}{Nsh} \right)^2 + \left(\frac{1}{\alpha_w}.I_{\delta,j,k}\right)^2 \right] }_{=:X} + \underbrace{ R_{strd} \times \sum_{k=1}^{N_h}
        2 \times \frac{I_{bundle,k}}{\alpha_w.Nsh} . \sum_{j=1}^{Nsh} I_{\delta,j,k} }_{=:Y} \\
    \end{aligned}
    \label{P_CC_1}
\end{equation}

The term $Y$ can be simplified, by substituting with (\ref{I_strand_component}), which gives:
\begin{equation}
\begin{aligned}
    Y = {} & R_{strd} \times \sum_{k=1}^{N_h}
        2 \times \frac{I_{bundle,k}}{\alpha_w.Nsh} . \sum_{j=1}^{Nsh} I_{\delta,j,k} \\
        = {} & R_{strd} \times \sum_{k=1}^{N_h}
        2 \times \frac{I_{bundle,k}}{\alpha_w.Nsh} . \frac{-j.k.\omega}{Z_{strd,act,k}}. \sum_{j=1}^{Nsh} \left[ \myMatrix{\phi_{A,k}}(j)-\frac{1}{Nsh}.\sum_{n=1}^{Nsh}\myMatrix{\phi_{A,k}}(n)\right] \\
        = {} & R_{strd} \times \sum_{k=1}^{N_h}
        2 \times \frac{I_{bundle,k}}{\alpha_w.Nsh} . \frac{-j.k.\omega}{Z_{strd,act,k}}. \left[ \sum_{j=1}^{Nsh} \myMatrix{\phi_{A,k}}(j) - \sum_{j=1}^{Nsh} \frac{1}{Nsh}.\sum_{n=1}^{Nsh}\myMatrix{\phi_{A,k}}(n) \right] \\
        = {} & R_{strd} \times \sum_{k=1}^{N_h}
        2 \times \frac{I_{bundle,k}}{\alpha_w.Nsh} . \frac{-j.k.\omega}{Z_{strd,act,k}}. \underbrace{ \left[ \sum_{j=1}^{Nsh} \myMatrix{\phi_{A,k}}(j) - \sum_{n=1}^{Nsh}\myMatrix{\phi_{A,k}}(n) \right] }_{=0} { } = { } 0
\end{aligned}
\end{equation}

Hence, $Y = 0$ and (\ref{P_CC_1}) becomes:

\begin{equation}
\begin{aligned}
    P_{CC} = {} & R_{strd} \times \sum_{k=1}^{N_h}       \sum_{j=1}^{Nsh} \left[ \left(\frac{I_{bundle,k}}{Nsh} \right)^2 + \left(\frac{1}{\alpha_w}.I_{\delta,j,k}\right)^2 \right] \\
    = {} & \underbrace{R_{strd} \times \sum_{k=1}^{N_h}       \sum_{j=1}^{Nsh}  \left(\frac{I_{bundle,k}}{Nsh} \right)^2}_{P_{CC=0}} + \underbrace{\frac{1}{\alpha_w^2} . \underbrace{R_{strd} \times \sum_{k=1}^{N_h}       \sum_{j=1}^{Nsh} \left(I_{\delta,j,k}\right)^2}_{P_{\delta CC, active}}}_{P_{\delta CC}}
\end{aligned}
\end{equation}

As defined in Section 2, the term $P_{CC=0}$ represents the losses occurring without the effect of circulating currents, and the terms $P_{\delta CC}$ and $P_{\delta CC,active}$ correspond to additional losses due to the effect of circulating currents with and without consideration of the end windings, respectively. Losses due to circulating currents $P_{CC}$ can then be simply expressed as follows:

\begin{equation}
\begin{aligned}
    P_{CC} = {} & P_{CC=0} + P_{\delta CC}\\
    P_{CC} = {} & P_{CC=0} + \frac{1}{\alpha_w^2} . P_{\delta CC, active}
\end{aligned}
\label{P_CC_2}
\end{equation}

The terms $P_{\delta CC,active}$, $P_{\delta CC}$, and  $\alpha_w$ are non-negative, hence:
\begin{equation}
    P_{CC} \ge P_{CC=0}
    \label{FProperty}
\end{equation}

This inequality (\ref{FProperty}) is demonstrated mathematically in the general case in \cite{arxiv_taha_2024_1}. Finally, it can be noticed from (\ref{P_CC_2}) that when the ratio $\alpha_w$ increases, the term $P_{\delta CC}$ decreases at a rate proportional to $\frac{1}{\alpha_w^2}$, and consequently the losses due to circulating currents $P_{CC}$ decrease at the same rate with the lower limit being the losses occurring in the bundle without the effect of circulating currents $P_{CC=0}$. For very high value of $\alpha_w$, the second term $P_{\delta CC}$ at the right hand side of (\ref{P_CC_2}) can be neglected and only the first term $P_{CC=0}$ remains, therefore the effect of circulating currents can be neglected.

\textbf{Summary:}\\
It was demonstrated that, in the case of a diagonal impedance matrix, increasing the ratio $\alpha_w$ leads to reducing losses due to circulating currents at a rate proportional to $\frac{1}{\alpha_w^2}$, hence, proportional to the inverse square of the end windings length. Circulating currents can be neglected for high value of $\alpha_w$ (equivalent to long end windings). In the cases $1$ and $2$ presented above, the impedance matrix has negligible off-diagonal entries compared to the diagonal entries. The property can be generalized to these cases because of the continuity of the inverse of a matrix with respect to its entries.

\end{proof}

\section{Case Application: Electric Machine}

\begin{figure}[!h]
\centering
\begin{subfigure}{.49\textwidth}
    \centering
    \includegraphics[clip,trim={0cm} {0cm} {1.71cm} {0.6cm},width=.9\linewidth]{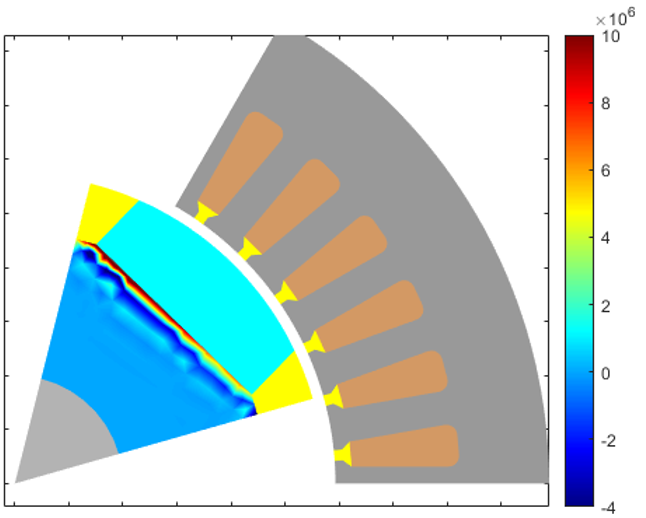}
    \caption{}
    \label{EM}
\end{subfigure}%
\begin{subfigure}{.5\textwidth}
    \centering
    \includegraphics[width=.8\textwidth]{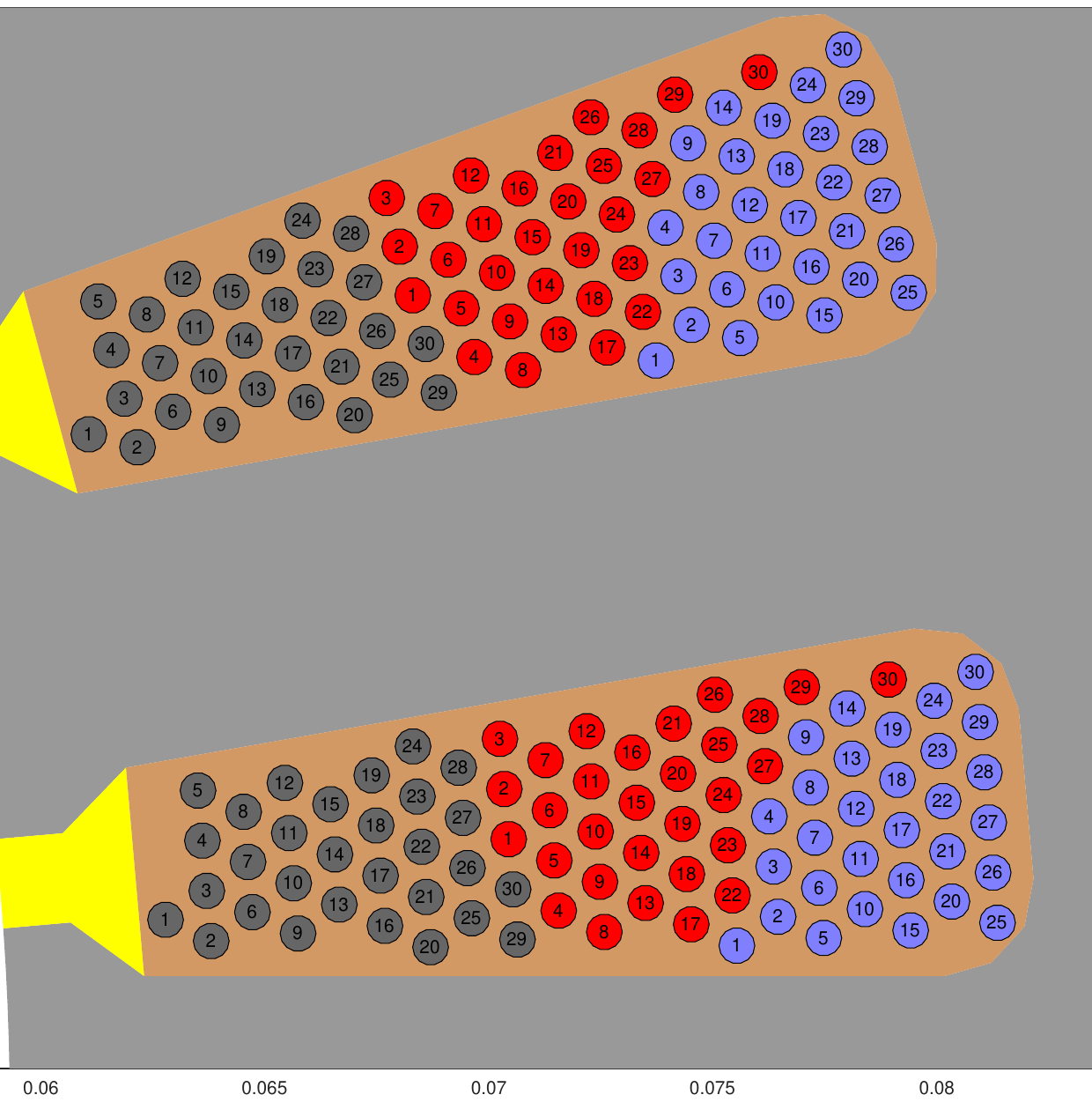}
    \vspace{-4.3em}
    \caption{}
    \label{SL}
\end{subfigure}
\caption{(a) S-PMSM \cite{arxiv_taha_2024_1} and (b) Slot Layout with specified locations of the strands \cite{arxiv_taha_2024_1}}
\end{figure}

Figure \ref{EM} presents a Surface Permanent Magnet Synchronous Machine (S-PMSM) which is the same machine used in previous authors' work \cite{arxiv_taha_2024_1} to evaluate circulating currents. This machine is used in this section to provide a numerical validation of the property mentioned above. The rotor has $6$ poles and the stator has $36$ slots with a distributed winding, allocated to the three phases ($2$ slots per pole and per phase). Each phase (A, B, C) is supplied with a sinusoidal waveform current. The slot layout \cite{arxiv_taha_2024_1} shown in Figure \ref{SL} illustrates the placements of the $30$ parallel-connected strands, enumerated from $1$ to $30$, with $3$ turns per slot, each represented by a different color (blue, red, and black). All the stator's slots have the same layout i.e. the same strands' locations. A co-simulation (finite element method coupled with circuit analysis) is conducted using the software EMDTool \cite{EMDTool} to evaluate the currents flowing through the $30$ parallel-connected strands of each phase. Three cases are studied: $\alpha_w=2$, $\alpha_w=2.5$, and $\alpha_w=3$.

\begin{figure}[!h]
\centering
\begin{subfigure}{.49\textwidth}
    \centering
    \includegraphics[width=.99\textwidth]{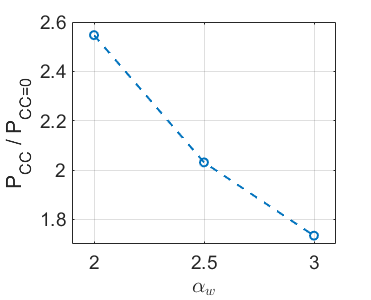}
    \caption{}
    \label{CCLoss_alpha}
\end{subfigure}%
\begin{subfigure}{.5\textwidth}
    \centering
    \includegraphics[width=.99\textwidth]{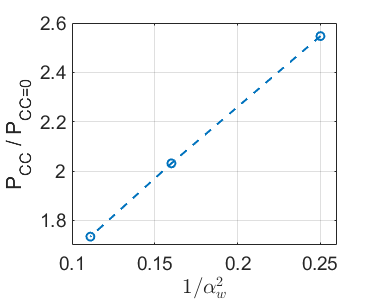}
    \caption{}
    \label{CCLoss_invalpha}
\end{subfigure}
\caption{Losses due to circulating currents with respect to (a) $\alpha_w$ and (b) $\frac{1}{\alpha_w^2}$}
\end{figure}

\begin{figure}[ht]
    \centering
    \begin{subfigure}[b]{0.49\textwidth}
        \centering
        \includegraphics[width=0.9\textwidth]{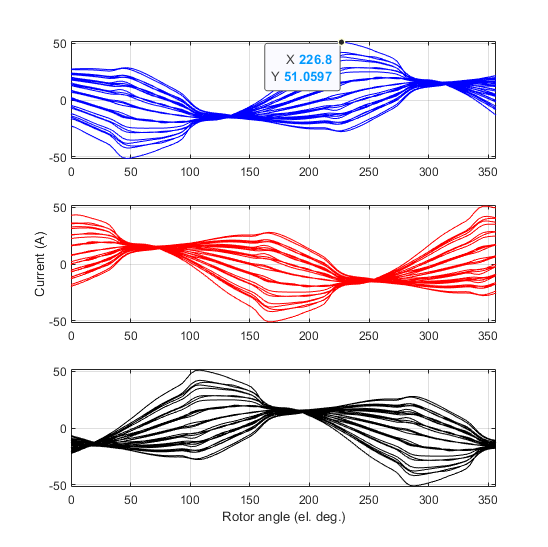} 
        \caption{$\alpha_w=2$}
        \label{CW_10}
    \end{subfigure}
    \begin{subfigure}[b]{0.49\textwidth}
        \centering
        \includegraphics[width=0.9\textwidth]{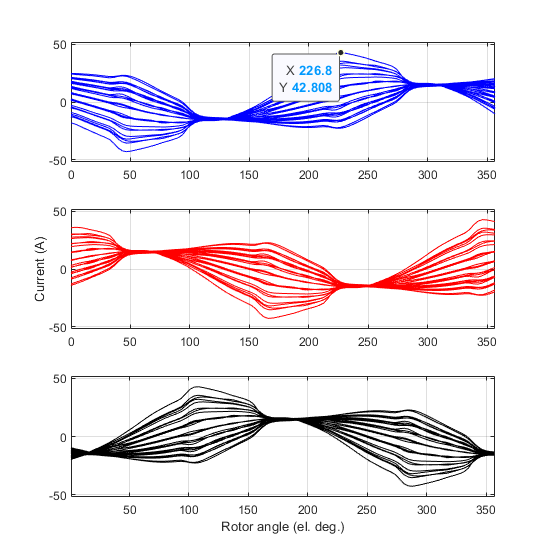}
        \caption{$\alpha_w=2.5$}
        \label{CW_15}
    \end{subfigure}
    \begin{subfigure}[b]{0.5\textwidth}
        \centering
        \includegraphics[width=0.9\textwidth]{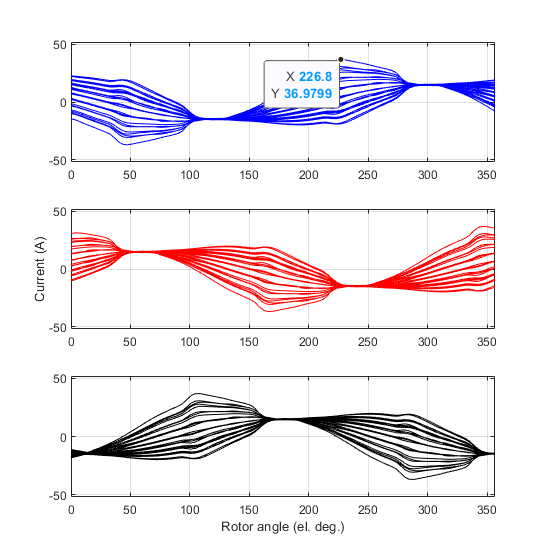}
        \caption{$\alpha_w=3$}
        \label{CW_20}
    \end{subfigure}
    \caption{Waveforms of the currents in the parallel strands for each phase for (a) $\alpha_w=2$, (b) $\alpha_w=2.5$, (c) $\alpha_w=3$ }
    \label{fig:test}
\end{figure}

Figure \ref{CCLoss_alpha} and Figure \ref{CCLoss_invalpha} show the losses due to circulating currents ($P_{CC}$ is normalized with $P_{CC=0}$, which are the losses without the effect of circulating currents) with respect to $\alpha_w$ and $\frac{1}{\alpha_w^2}$, respectively. It can be observed from Figure \ref{CCLoss_alpha} that as $\alpha_w$ increases, the losses caused by circulating currents decrease. Further, there is a linear relation between circulating currents losses and $\frac{1}{\alpha_w^2}$ as shown in Figure \ref{CCLoss_invalpha}. This means that circulating currents losses decrease at a rate proportional to $\frac{1}{\alpha_w^2}$, which is consistent with the property above. The instantaneous current flowing in each strand for the three phases is represented in Figure \ref{CW_10}, Figure \ref{CW_15}, and Figure \ref{CW_20} for the three cases $\alpha_w=2$, $\alpha_w=2.5$, and $\alpha_w=3$, respectively. The currents flowing in the parallel strands have non-sinusoidal waveforms, with different magnitude and phase due to the effect of circulating currents. The total current in the bundle is not equally shared, which leads to additional losses due to circulating currents \cite{arxiv_taha_2024_1}. Ideally, when circulating currents do not occur, the current is evenly shared with a sinusoidal waveform in each strand \cite{arxiv_taha_2024_1}. The same results are noticed for the three phases. Further, when the ratio $\alpha_w$ increases ($2 \xrightarrow[]{} 2.5 \xrightarrow[]{} 3$), the maximum current value in the parallel strands decreases ($51.05 \xrightarrow[]{} 42.8 \xrightarrow[]{} 36.9$, respectively). This implies that with longer end windings, the uneven distribution of current becomes less pronounced, leading toward a more uniform current sharing.

\section{Conclusion}

In this paper, we presented the positive impact of the end windings length in reducing the losses due to circulating currents and allowing a more uniform sharing of the current between the parallel strands. Generated losses due to circulating currents decrease at a rate proportional to the inverse square of the end windings length. This property was proven mathematically using the hybrid model for circulating currents in the case where the impedance matrix has negligible off-diagonal entries compared to the diagonal entries, extending the authors' previous work and framing it within a rigorous mathematical context. A case study of an electric machine was presented to study the impact of the end windings length on losses due to circulating currents and the results obtained from finite element analysis are coherent with this property.

\section*{Acknowledgment}
This research is funded by the Research Council of Finland CoE HiECSs under Grant 346438.


\bibliographystyle{IEEEtran}

\bibliography{Doc_CC_EW}

\begin{thebibliography}{10}
\providecommand{\url}[1]{#1}
\csname url@samestyle\endcsname
\providecommand{\newblock}{\relax}
\providecommand{\bibinfo}[2]{#2}
\providecommand{\BIBentrySTDinterwordspacing}{\spaceskip=0pt\relax}
\providecommand{\BIBentryALTinterwordstretchfactor}{4}
\providecommand{\BIBentryALTinterwordspacing}{\spaceskip=\fontdimen2\font plus
\BIBentryALTinterwordstretchfactor\fontdimen3\font minus \fontdimen4\font\relax}
\providecommand{\BIBforeignlanguage}[2]{{%
\expandafter\ifx\csname l@#1\endcsname\relax
\typeout{** WARNING: IEEEtran.bst: No hyphenation pattern has been}%
\typeout{** loaded for the language `#1'. Using the pattern for}%
\typeout{** the default language instead.}%
\else
\language=\csname l@#1\endcsname
\fi
#2}}
\providecommand{\BIBdecl}{\relax}
\BIBdecl

\bibitem{arxiv_taha_2024_1}
\BIBentryALTinterwordspacing
T.~E. Hajji, A.~Lehikoinen, and A.~Belahcen, ``Circulating currents in windings: Fundamental property,'' 2024. [Online]. Available: \url{https://arxiv.org/abs/2410.12748}
\BIBentrySTDinterwordspacing

\bibitem{Access_Taha_2024}
T.~El~Hajji, S.~Hlioui, F.~Louf, M.~Gabsi, A.~Belahcen, G.~Mermaz-Rollet, and M.~Belhadi, ``Ac losses in windings: Review and comparison of models with application in electric machines,'' \emph{IEEE Access}, vol.~12, pp. 1552--1569, 2024.

\bibitem{Machines_Taha_2023}
T.~El~Hajji, S.~Hlioui, F.~Louf, M.~Gabsi, G.~Mermaz-Rollet, and M.~Belhadi, ``Optimal design of high-speed electric machines for electric vehicles: A case study of 100 kw v-shaped interior pmsm,'' \emph{Machines}, vol.~11, no.~1, p.~57, Jan 2023.

\bibitem{Aerospace_Taha_2024}
T.~El~Hajji, A.~Hemeida, A.~Lehikoinen, F.~Martin, and A.~Belahcen, ``Optimal design of high specific power electric machines for fully electric regional aircraft: A case study of 1mw s-pmsm,'' \emph{Aerospace}, vol.~11, no.~10, 2024.

\bibitem{HighSpeed_1CN}
J.~Zhang, H.~Jiang, Z.~Zhang, L.~Yu, and X.~Zhu, ``Ac loss analytic method and optimization of litz winding for high-speed electrical machines,'' \emph{IEEE Transactions on Industrial Electronics}, vol.~71, no.~4, pp. 3330--3341, 2024.

\bibitem{ICEM_Taha_2024}
T.~El~Hajji, A.~Lehikoinen, A.~Hemeida, and A.~Belahcen, ``Optimal design of cost-effective e-machines for traction: A case study of 150kw v-shaped pmsm,'' in \emph{2024 International Conference on Electrical Machines (ICEM)}, 2024, pp. 1--5.

\bibitem{PhD_Taha_2023}
\BIBentryALTinterwordspacing
T.~El~Hajji, ``Modélisation et optimisation de machines électriques à haute vitesse pour les véhicules électriques,'' Ph.D. dissertation, Electrical Engineering, Université Paris-Saclay, 2023, 2023UPAST017. [Online]. Available: \url{http://www.theses.fr/2023UPAST017}
\BIBentrySTDinterwordspacing

\bibitem{Transformer_1}
A.~Reatti and M.~Kazimierczuk, ``Comparison of various methods for calculating the ac resistance of inductors,'' \emph{IEEE Transactions on Magnetics}, vol.~38, no.~3, pp. 1512--1518, 2002.

\bibitem{Transformer_2}
N.~Rasekh, J.~Wang, and X.~Yuan, ``In-situ measurement and investigation of winding loss in high-frequency cored transformers under large-signal condition,'' \emph{IEEE Open Journal of Industry Applications}, vol.~3, pp. 164--177, 2022.

\bibitem{Transformer_4}
D.~Elizondo, E.~L. Barrios, A.~Ursúa, and P.~Sanchis, ``Analytical modeling of high-frequency winding loss in round-wire toroidal inductors,'' \emph{IEEE Transactions on Industrial Electronics}, vol.~70, no.~6, pp. 5581--5591, 2023.

\bibitem{Transformer_5}
T.~Luo, M.~G. Niasar, and P.~Vaessen, ``2-d winding losses calculation for round conductor coil,'' \emph{IEEE Transactions on Power Electronics}, vol.~38, no.~4, pp. 5107--5117, 2023.

\bibitem{Transformer_6}
C.~Peng, G.~Chen, B.~Wang, and J.~Song, ``Semi-analytical ac resistance prediction model for litz wire winding in high-frequency transformer,'' \emph{IEEE Transactions on Power Electronics}, vol.~38, no.~10, pp. 12\,730--12\,742, 2023.

\bibitem{Transformer_7}
A.~Arruti, I.~Aizpuru, M.~Mazuela, Z.~Ouyang, and M.~A. Andersen, ``Evolution of classical 1-d-based models and improved approach for the characterization of litz wire losses,'' \emph{IEEE Transactions on Power Electronics}, vol.~39, no.~12, pp. 16\,371--16\,381, 2024.

\bibitem{Placement_Random_1_RefFI_Antti}
A.~Lehikoinen, N.~Chiodetto, E.~Lantto, A.~Arkkio, and A.~Belahcen, ``Monte carlo analysis of circulating currents in random-wound electrical machines,'' \emph{IEEE Transactions on Magnetics}, vol.~52, no.~8, pp. 1--12, 2016.

\bibitem{Placement_Random_2_RefFI_Antti}
A.~Lehikoinen, N.~Chiodetto, A.~Arkkio, and A.~Belahcen, ``Improved sampling algorithm for stochastic modelling of random-wound electrical machines,'' \emph{The Journal of Engineering}, vol. 2019, no.~17, pp. 3976--3980, 2019.

\bibitem{Placement_Random_3_PhD_Antti_2017}
\BIBentryALTinterwordspacing
A.~Lehikoinen, ``Circulating and eddy-current losses in random-wound electrical machines,'' Ph.D. dissertation, Department of Electrical Engineering and Automation, Aalto University, 2017. [Online]. Available: \url{https://aaltodoc.aalto.fi/items/5f7ca4cf-61a1-4038-99ed-be08db2c813c}
\BIBentrySTDinterwordspacing

\bibitem{Placement_Precise_2_RefFI_PhD_Jahirul_2010}
\BIBentryALTinterwordspacing
M.~J. Islam, ``Finite-element analysis of eddy currents in the form-wound multi-conductor windings of electrical machines,'' Ph.D. dissertation, Department of Electrical Engineering, Aalto University, 2010. [Online]. Available: \url{https://aaltodoc.aalto.fi/items/77a78e68-4fdc-46c5-99ff-9fe73d8807f0}
\BIBentrySTDinterwordspacing

\bibitem{Asymmetry_1}
S.~Mukundan, H.~Dhulipati, Z.~Li, M.~S. Toulabi, J.~Tjong, and N.~C. Kar, ``Coupled magnetic circuit-based design of an ipmsm for reduction of circulating currents in asymmetrical star–delta windings,'' \emph{IEEE Transactions on Transportation Electrification}, vol.~8, no.~2, pp. 2971--2984, 2022.

\bibitem{Abnormality_1}
Y.~Chulaee, D.~Lewis, A.~Mohammadi, G.~Heins, D.~Patterson, and D.~M. Ionel, ``Circulating and eddy current losses in coreless axial flux pm machine stators with pcb windings,'' \emph{IEEE Transactions on Industry Applications}, vol.~59, no.~4, pp. 4010--4020, 2023.

\bibitem{pramodCC}
\BIBentryALTinterwordspacing
P.~Pramod, ``Circulating current induced electromagnetic torque generation in electric machines with delta windings,'' 2023. [Online]. Available: \url{https://arxiv.org/abs/2310.06469}
\BIBentrySTDinterwordspacing

\bibitem{ICEM_Taha_2020}
T.~El~Hajji, S.~Hlioui, F.~Louf, M.~Gabsi, G.~Mermaz-Rollet, and M.~Belhadi, ``Hybrid model for ac losses in high speed pmsm for arbitrary flux density waveforms,'' in \emph{2020 International Conference on Electrical Machines (ICEM)}, vol.~1, 2020, pp. 2426--2432.

\bibitem{SkinProx_1_UKSE}
D.~Y. Um, R.~Kumar, T.~Batra, L.~Sjöberg, and G.~Atkinson, ``A comparative study of ac winding loss calculation for axial flux permanent magnet machines based on finite element analysis,'' in \emph{2024 International Conference on Electrical Machines (ICEM)}, 2024, pp. 1--7.

\bibitem{SkinProx_2_USA_Ayman}
F.~Wu, A.~M. EL-Refaie, and A.~Al-Qarni, ``Minimization of winding ac losses using inhomogeneous electrical conductivity enabled by additive manufacturing,'' \emph{IEEE Transactions on Industry Applications}, vol.~58, no.~3, pp. 3447--3458, 2022.

\bibitem{SkinProx_3_DE}
Y.~Ma, K.~Reutlinger, A.~Stirban, and M.~Doppelbauer, ``Extension of analytical methods for ac copper loss estimation in high-speed axial flux machines with concentrated flat wire coils,'' in \emph{2024 International Conference on Electrical Machines (ICEM)}, 2024, pp. 1--8.

\bibitem{SkinProx_4_UK}
Y.~Yuan, M.~A. Darmani, Y.~Bao, X.~Zhang, D.~Gerada, and H.~Zhang, ``Analysis of proximity loss of electrical machines using mesh-based magnetic equivalent circuit,'' \emph{IEEE Transactions on Transportation Electrification}, pp. 1--1, 2024.

\bibitem{Transposition_1_Ref22}
P.~B. Reddy, T.~M. Jahns, and T.~P. Bohn, ``Transposition effects on bundle proximity losses in high-speed pm machines,'' in \emph{2009 IEEE Energy Conversion Congress and Exposition}, 2009, pp. 1919--1926.

\bibitem{Transposition_2_Ref23}
P.~B. Reddy and T.~M. Jahns, ``Analysis of bundle losses in high speed machines,'' in \emph{The 2010 International Power Electronics Conference - ECCE ASIA -}, 2010, pp. 2181--2188.

\bibitem{Transposition_3_RefCN1}
K.~Zhang, Y.~Liang, P.~Yang, and D.~Wang, ``Ac loss calculation of integral transposed winding in pm motor for electric vehicles,'' \emph{IEEE Transactions on Transportation Electrification}, pp. 1--1, 2024.

\bibitem{Transposition_4_RefCN2}
X.~Ju, Y.~Cheng, B.~Du, M.~Yang, D.~Yang, and S.~Cui, ``Ac loss analysis and measurement of a hybrid transposed hairpin winding for ev traction machines,'' \emph{IEEE Transactions on Industrial Electronics}, vol.~70, no.~4, pp. 3525--3536, 2023.

\bibitem{Transposition_5_RefNW}
F.~Maurer and J.~K. Nøland, ``A rectangular end-winding model for enhanced circulating current prediction in ac machines,'' \emph{IEEE Transactions on Energy Conversion}, vol.~36, no.~1, pp. 291--299, 2021.

\bibitem{Transposition_6_RefUK}
J.~Hoole, P.~H. Mellor, and N.~Simpson, ``Towards data-driven strand transposition simulation of multistrand random windings,'' in \emph{2024 International Conference on Electrical Machines (ICEM)}, 2024, pp. 1--7.

\bibitem{Transposition_7_RefJP}
T.~Kono, J.~Asama, and H.~Shida, ``Reduction of circulating current by bundle inversion in electrical machines with distributed winding,'' in \emph{2023 IEEE Energy Conversion Congress and Exposition (ECCE)}, 2023, pp. 3818--3821.

\bibitem{Transposition_8_RefCN}
Y.~Jiang, J.~Chen, H.~Wang, and D.~Wang, ``Semi-analytical method of form-wound winding loss considering circulating current effect,'' \emph{IEEE Transactions on Magnetics}, vol.~58, no.~2, pp. 1--6, 2022.

\bibitem{Circular_Wire_Ref18}
P.~Mellor, J.~Hoole, and N.~Simpson, ``Computationally efficient prediction of statistical variance in the ac losses of multi-stranded windings,'' in \emph{2021 IEEE Energy Conversion Congress and Exposition (ECCE)}, 2021, pp. 3887--3894.

\bibitem{Circular_Wire_Ref19}
J.~Hoole, P.~H. Mellor, and N.~Simpson, ``Designing for conductor lay and ac loss variability in multistrand stator windings,'' \emph{IEEE Transactions on Industry Applications}, vol.~59, no.~2, pp. 1394--1404, 2023.

\bibitem{Circular_Wire_Ref20}
J.~Hoole, P.~H. Mellor, N.~Simpson, and D.~North, ``Statistical simulation of conductor lay and ac losses in multi-strand stator windings,'' in \emph{2021 IEEE International Electric Machines \& Drives Conference (IEMDC)}, 2021, pp. 1--8.

\bibitem{Guideline_1}
G.~Berardi and N.~Bianchi, ``Design guideline of an ac hairpin winding,'' in \emph{2018 XIII International Conference on Electrical Machines (ICEM)}, 2018, pp. 2444--2450.

\bibitem{Guideline_2}
T.~Zou, D.~Gerada, A.~L. Rocca, M.~Moslemin, A.~Cairns, M.~Cui, A.~Bardalai, F.~Zhang, and C.~Gerada, ``A comprehensive design guideline of hairpin windings for high power density electric vehicle traction motors,'' \emph{IEEE Transactions on Transportation Electrification}, vol.~8, no.~3, pp. 3578--3593, 2022.

\bibitem{Rectangular_Wire_1}
D.~L. Bezerra, T.~Zou, A.~L. Rocca, M.~Cui, H.~Huang, J.~Al-Tayie, J.~Majer, A.~Cairns, and C.~Gerada, ``Analysis of circulating current in hairpin windings due to manufacturing deviation,'' in \emph{2023 IEEE International Electric Machines \& Drives Conference (IEMDC)}, 2023, pp. 1--7.

\bibitem{FEA_1_Antti}
A.~Lehikoinen and A.~Arkkio, ``Efficient finite-element computation of circulating currents in thin parallel strands,'' \emph{IEEE Transactions on Magnetics}, vol.~52, no.~3, pp. 1--4, 2016.

\bibitem{Hybrid_Ref33}
C.~Roth, F.~Birnkammer, and D.~Gerling, ``Analytical model for ac loss calculation applied to parallel conductors in electrical machines,'' in \emph{2018 XIII International Conference on Electrical Machines (ICEM)}, 2018, pp. 1088--1094.

\bibitem{EMDTool}
\BIBentryALTinterwordspacing
``Emdtool,'' accessed: 2024-10-20. [Online]. Available: \url{www.emdtool.com}
\BIBentrySTDinterwordspacing

\end{thebibliography}

\end{document}